\documentstyle[wrapfig,epsfig]{aipproc}
\def\us{$^\dagger$}

\def\fsize={2}
\def\LCDM{$\Lambda$CDM}
\def\dndE{\left({{\rm d}\phi\over {\rm d}E}\right)}

\begin{document}

\begin{minipage}[t]{.2\linewidth}
\leavevmode
 \hspace*{-.8cm}                       
\psfig{file=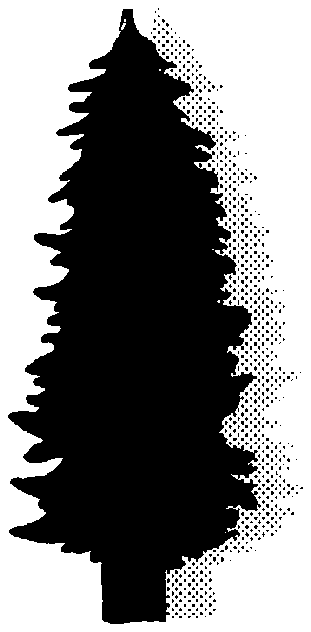,width=3cm}   
\end{minipage} \hfill
\begin{minipage}[b]{.45\linewidth}     
\rightline{SCIPP 99/43}                
\rightline{      1999}
\vspace{3.5cm}
\end{minipage}
\vskip3pc

\title{The Implications of Galaxy Formation Models for the TeV
   Observations of Current Detectors}
\author{L.M. Boone\us, J.S. Bullock$^\ddagger$, J.R. Primack\us, D.A. Williams\us}
\address{\us Santa Cruz Institute for Particle Physics, University of
   California, Santa Cruz, CA 95064\\
   $^\ddagger$Astronomy Department, Ohio State University, 
   Columbus, OH 43215}
\maketitle

\begin{abstract}
This paper represents a step toward constraining galaxy
formation models via TeV gamma ray observations.  We use 
semi-analytic models of galaxy formation to predict a spectral
distribution for the intergalactic infrared photon field, which
in turn yields information about the absorption of TeV gamma
rays from extra-galactic sources.  By making predictions for
integral flux observations at $>$200 GeV for several known EGRET
sources, 
we directly compare our models with current observational
upper limits obtained by Whipple.  In addition, our predictions
may offer a guide to the observing programs for the current
population of TeV
gamma ray observatories.
\end{abstract}

\section*{Introduction}

As shown previously \cite{macminn96,primack99}, measurements of
the extra-galactic background light (EBL) may be used to probe
models of galaxy formation.  The model predictions for the EBL can
be probed indirectly via the attenuation of high energy gamma rays,
due to pair production with the EBL photon field.  The $\gamma
\gamma\!\rightarrow\!e^+e^-$ cross section \cite{gould67} is
maximized when
$E_{\gamma}E_{EBL}\sim 2m_e^2$.
From this, we expect TeV gamma rays to be primarily absorbed by
EBL photons in the infrared (IR) region of the spectrum.  Galaxy
formation models
which differ in their predicted amount of infrared EBL should be
distinguishable by their predicted TeV gamma ray absorption.
Here we present results of simulations for the observed spectra of six
candidate blazars, including the absorption corrected integral flux
we would expect from two plausible galaxy formation models.
Other
work calculating EBL absorption of TeV spectra under somewhat different
assumptions is presented in \cite{stecker} and \cite{reshmi}.

\section*{Semi Analytic Models}

We have
modeled the EBL using the semi-analytic models (SAMs) of galaxy
formation discussed in \cite{somerville99}, and a similar \LCDM\
cosmology where $\Omega_{\Lambda}=.7$, $\Omega_m=0.3$, and $h=0.7$,
normalized such that the rms mass variance on the scale 8 Mpc/h is 1.

\begin{figure}
\centerline{
   \epsfig{file=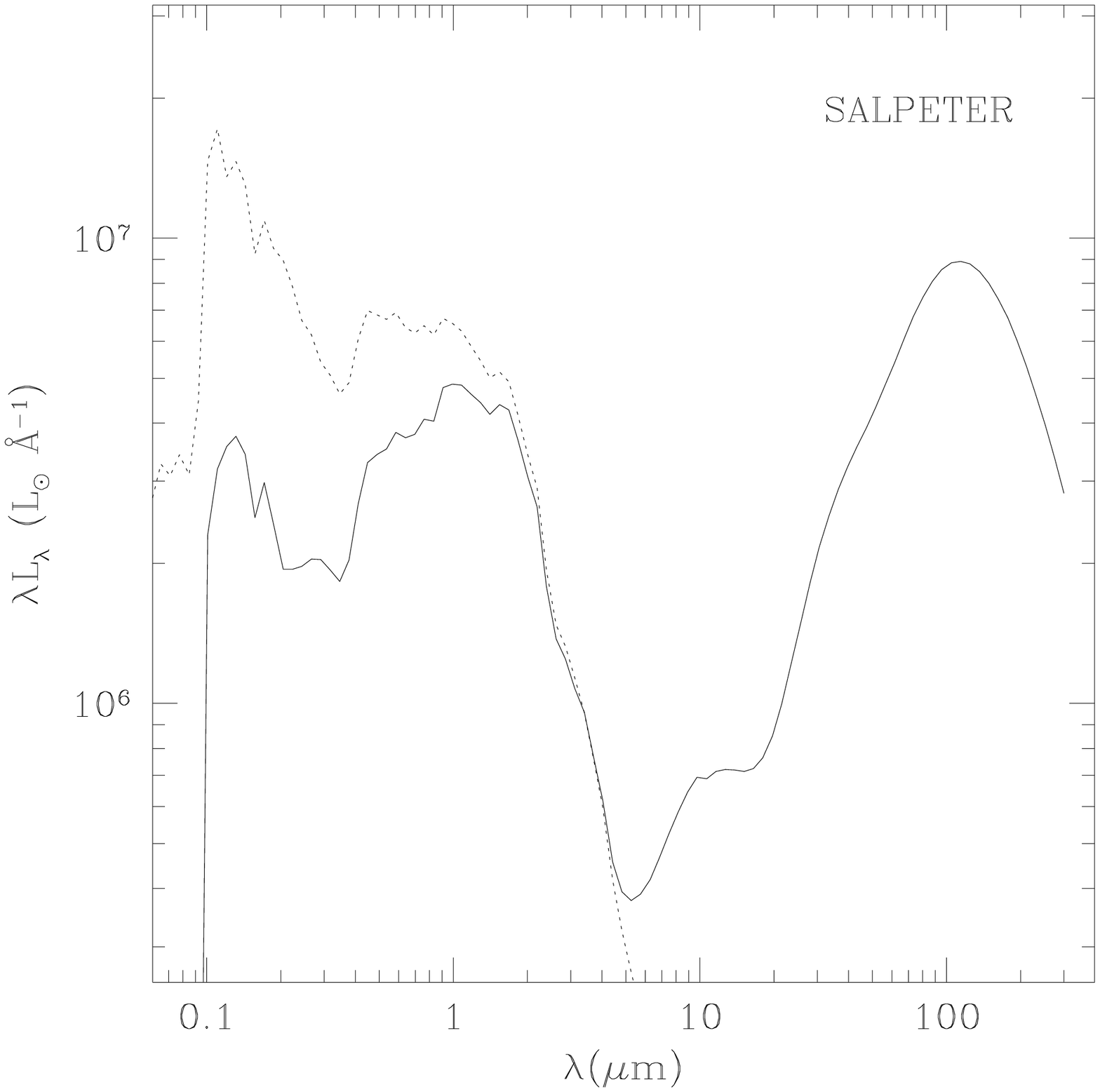,height=2.9in,width=2.9in}
   \epsfig{file=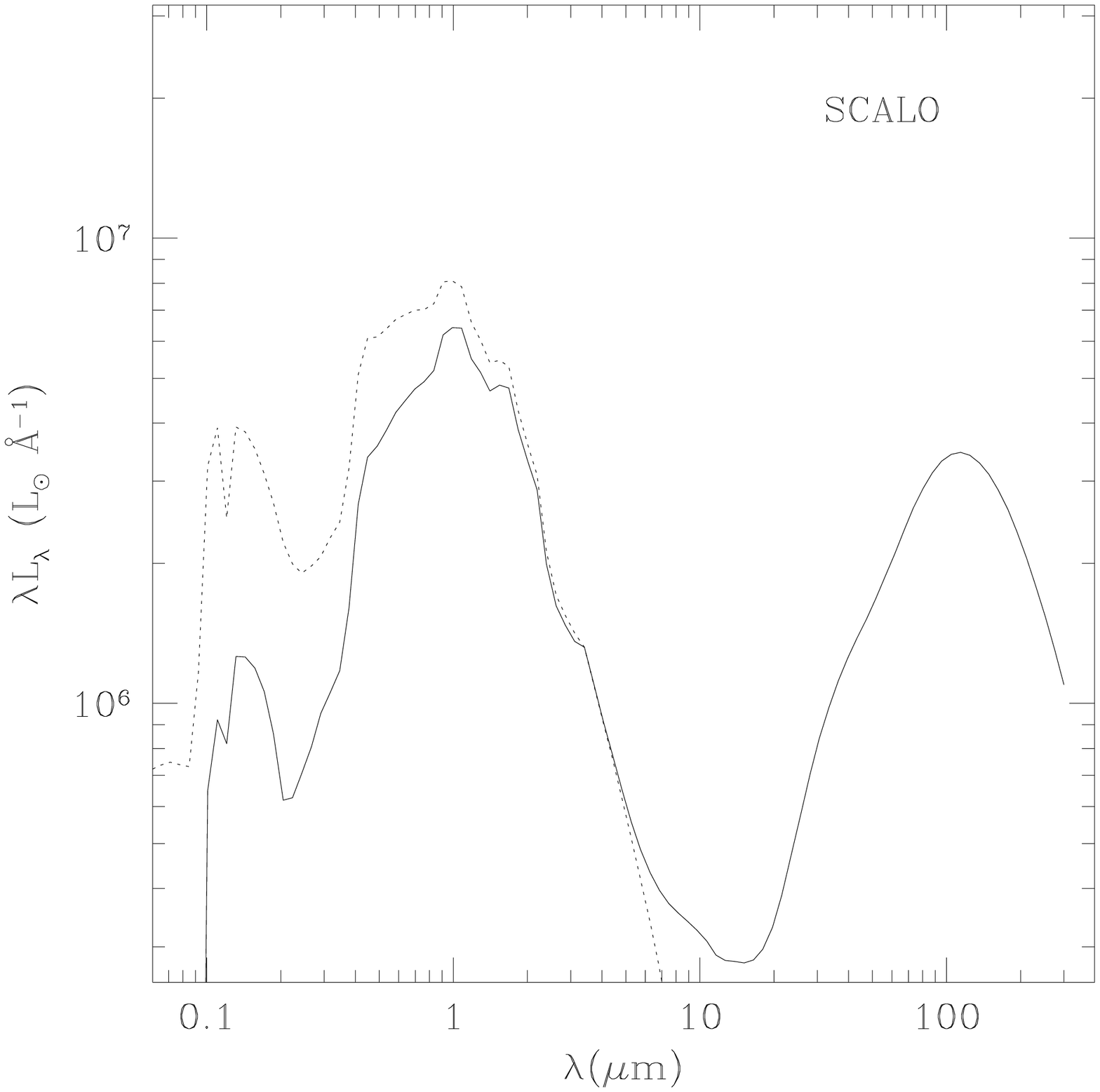,height=2.9in,width=2.9in}
   }
\caption{These figures represent the spectrum of an average ``Milky Way''
   sized galaxy for each
   of the two IMFs considered here.  The dotted lines indicate the
   starlight spectra without the effects of dust, while the solid lines
   represent the spectra with the effects of extinction and emission
   by dust.
   }
\label{figir}
\end{figure}

We have modeled the EBL for our \LCDM\ cosmology using two popular
models of the stellar initial mass function (IMF): the Salpeter IMF
\cite{salpeter} and the Scalo IMF \cite{scalo}.  The IMF, which
describes the stellar mass distribution, affects the wavelength
distribution of the starlight produced, and hence the wavelength
distribution of
the EBL.

The main difference between the two models is that the Salpeter IMF
has a larger fraction of high-mass stars than Scalo, provided both
are normalized to the same total mass of stars.  Sample spectra for
both models appear in Fig.~\ref{figir}.  Note that the Salpeter IMF
has more ultraviolet light than the corresponding Scalo model. This
is because high-mass stars produce more ultraviolet light than
low-mass stars.  Furthermore, because dust absorbs ultraviolet light
and emits around 100 \micron, the additional ultraviolet light
produced by the Salpeter IMF results in a greater amount of $\sim$100
\micron\ EBL.  Due to this enhanced 100 \micron\ bump, the Salpeter
IMF should produce stronger attenuation for corresponding gamma rays
between about 10 GeV and 10 TeV.

\section*{Implications for T\lowercase{e}V Astronomy}

\begin{figure}
\centerline{
   \epsfig{file=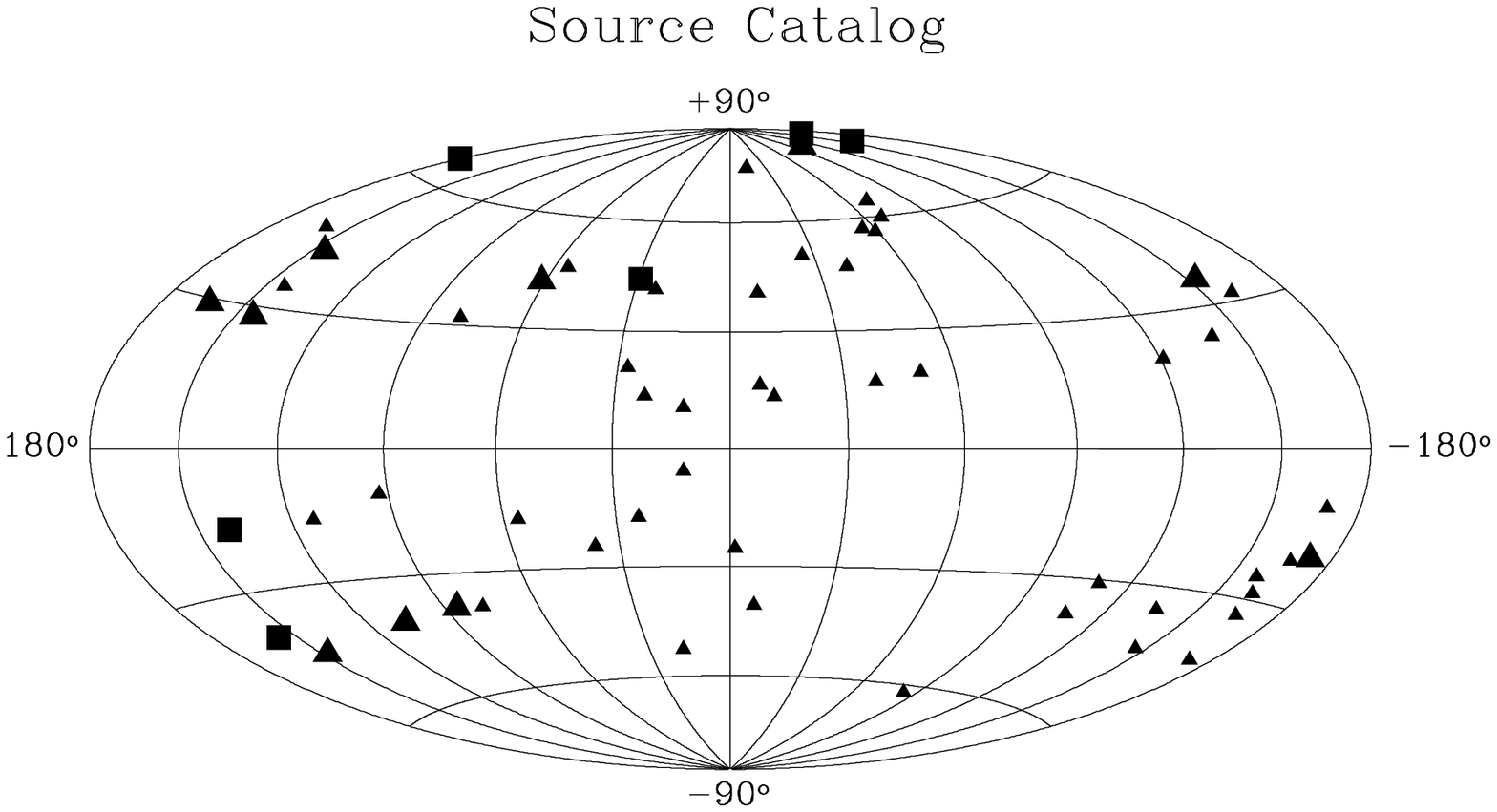,height=2.8in,width=2.8in}
   \epsfig{file=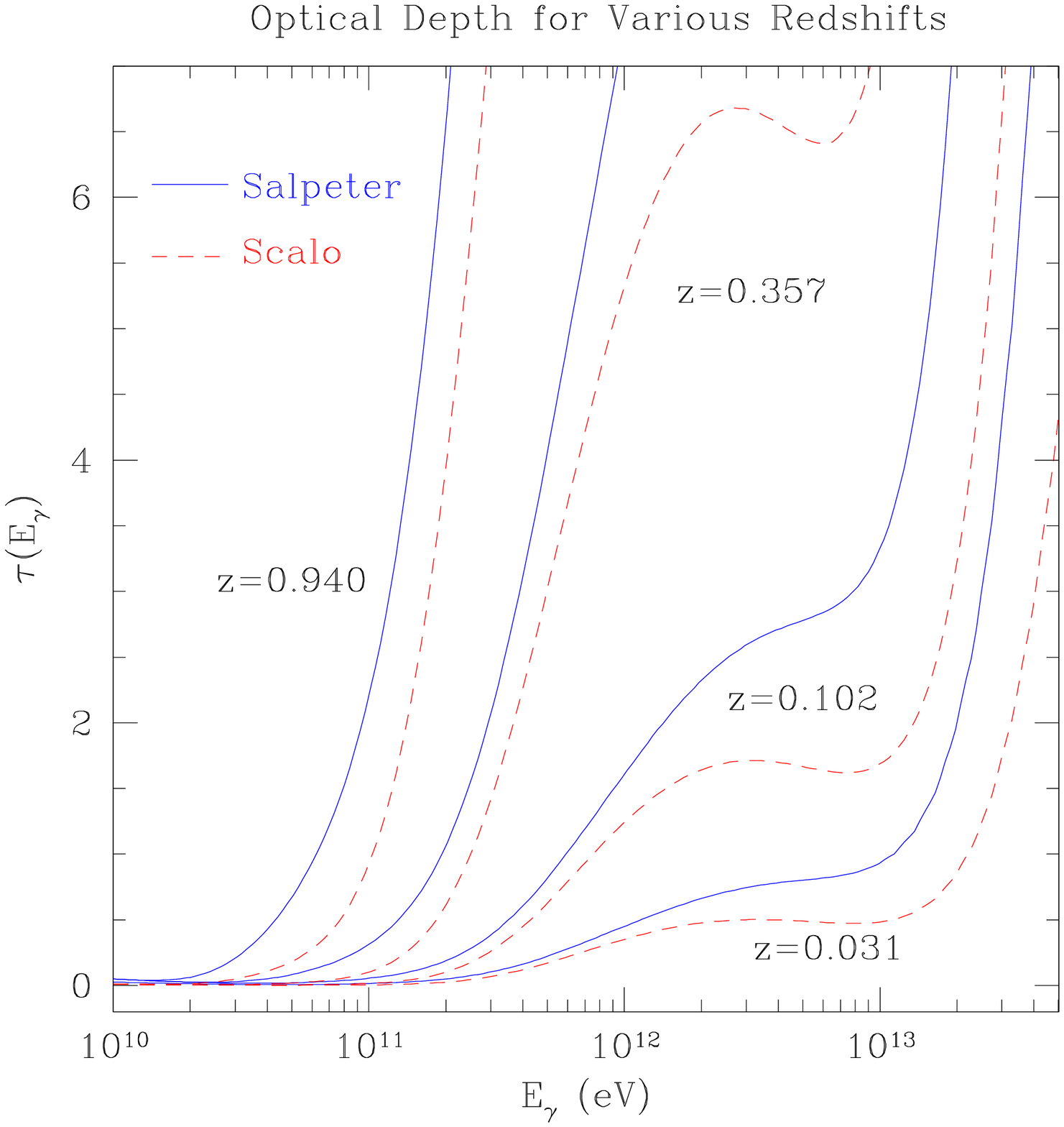,height=2.8in,width=2.8in}
   }
\caption{{\bf a)} Source catalog in galactic coordinates.  Small points
   denote all considered EGRET AGN, of which 16 (large points) were
   selected as ``candidate'' sources.  We have plotted absorbed spectra
   for six of these candidate sources (large squares) {\bf b)} Optical
   depth for selected redshifts as a function of gamma ray energy.}
\label{figtau}
\end{figure}

We chose our candidate TeV sources from the third EGRET catalog
\cite{hartman99}.
Of the 67 AGN listed in that catalog, we considered a subset of 60
sources for which there was complete data.
These sources are shown in
Fig.~\ref{figtau}a.
Of the 60 sources considered, 16 were chosen as candidate sources for
TeV observations based on the hardness of their spectra and their
integral flux above 100 MeV.  Candidate sources appear as large points
in Fig.~\ref{figtau}a.  We then simulated the observed integral flux
for six of these sources (large squares), both with and without
absorption corrections.  Of our six sources, one (Mrk421) is an X-ray
selected BL Lac, one (4C+29.45) is a flat spectrum radio quasar, and
the rest are radio selected BL Lacs.

\begin{figure}
   \epsfig{file=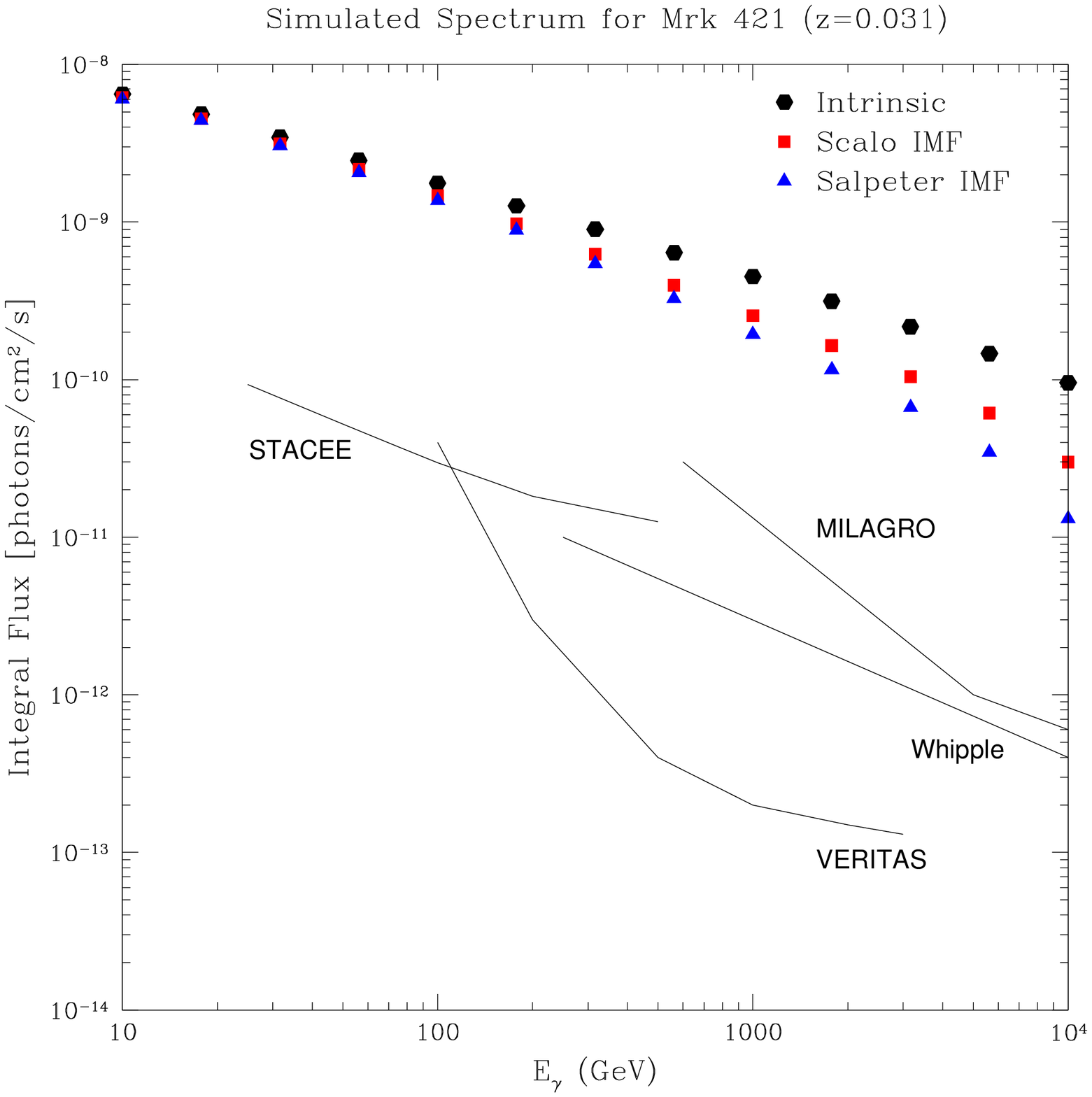,height=2.73in,width=2.73in}
   \hfill
   \epsfig{file=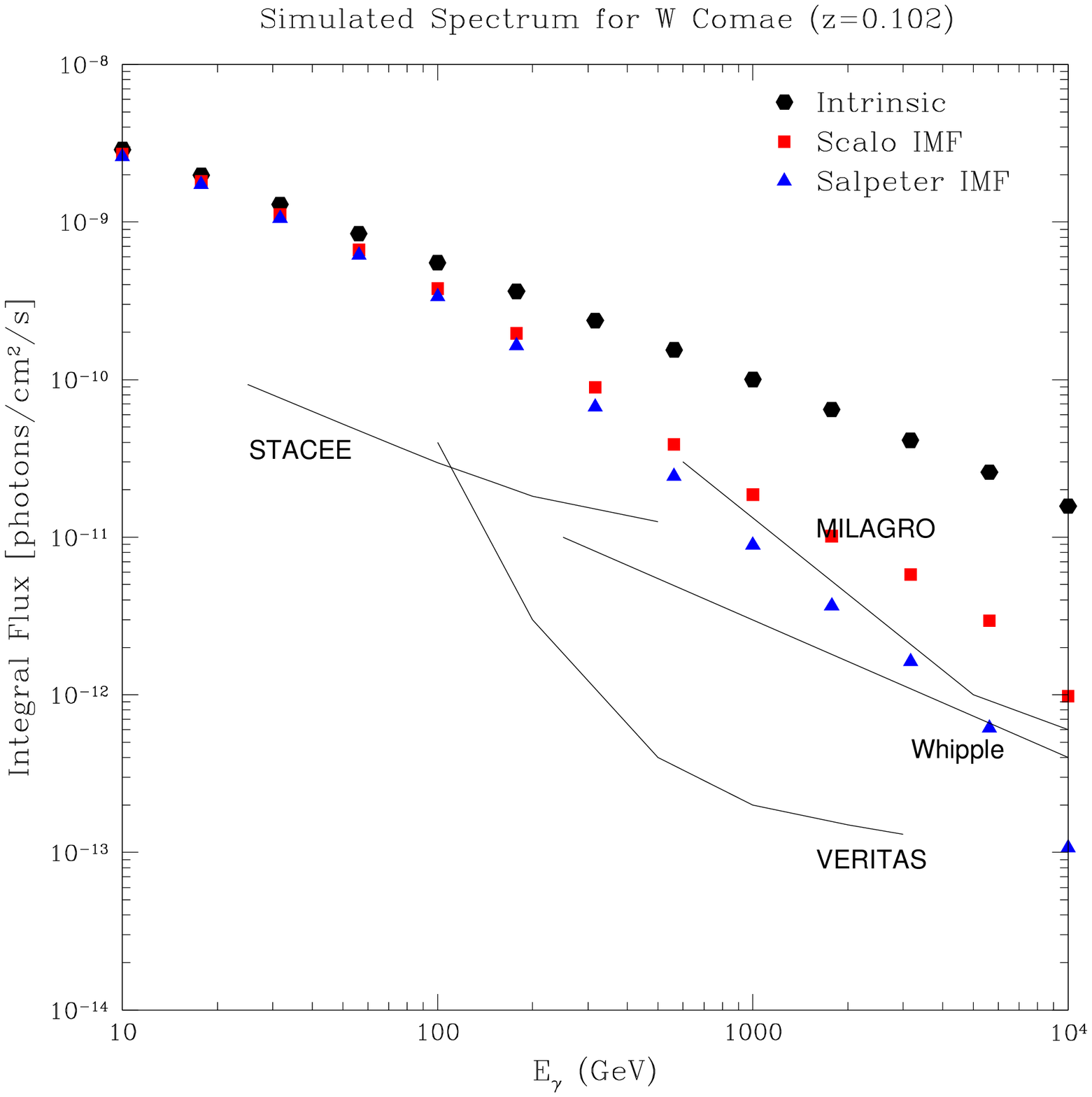,height=2.73in,width=2.73in}
   \epsfig{file=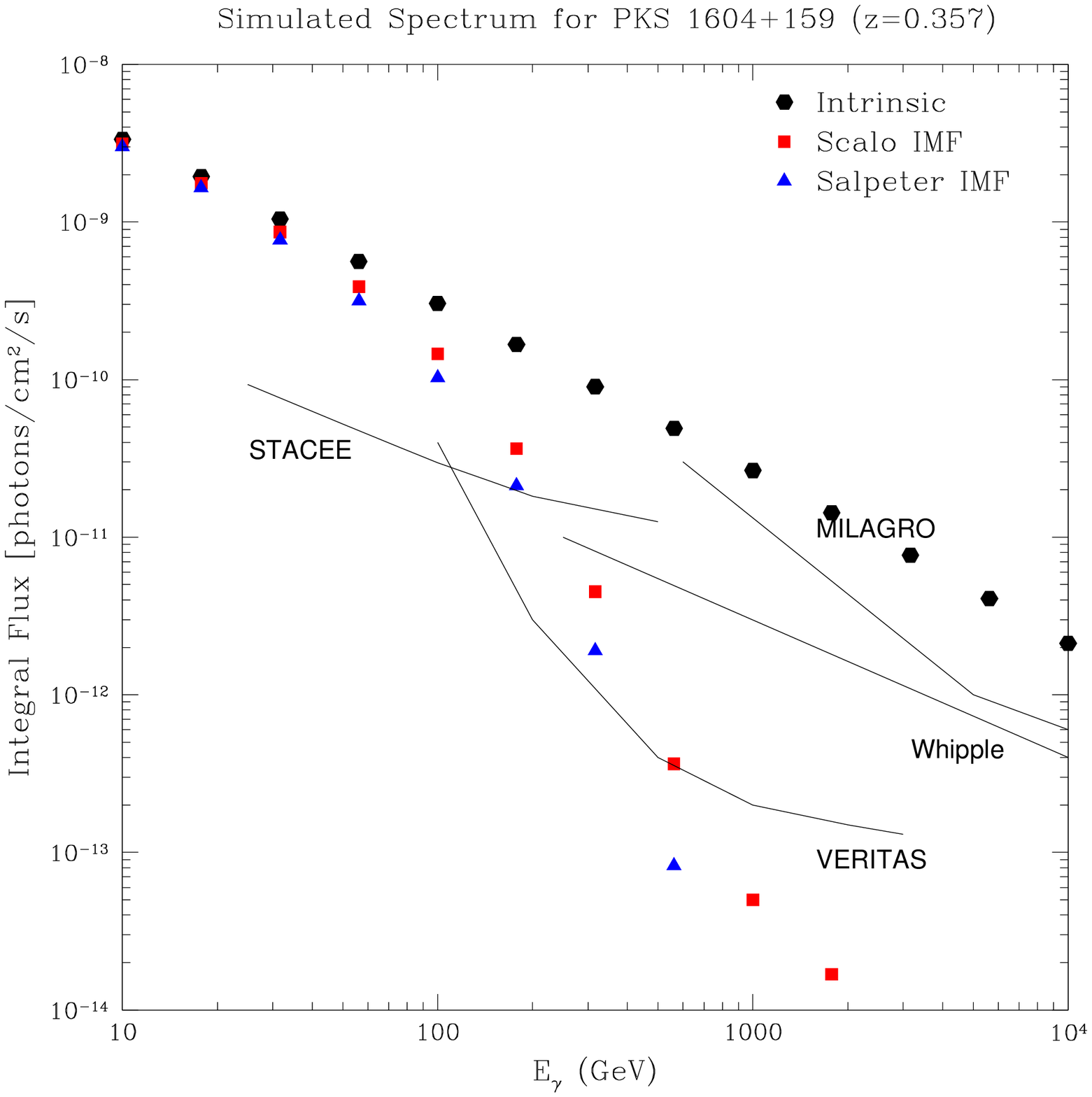,height=2.73in,width=2.73in}
   \hfill
   \epsfig{file=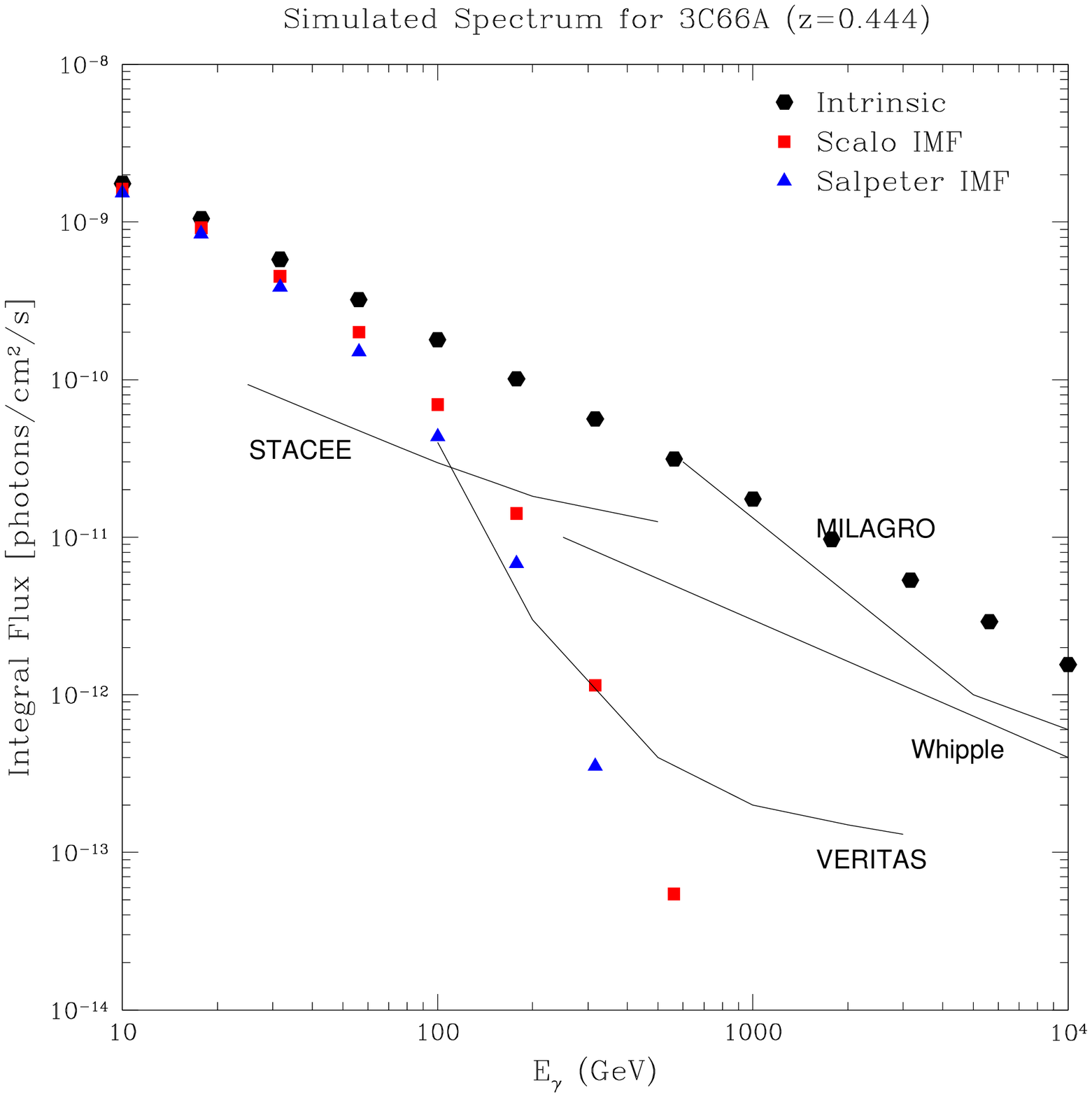,height=2.73in,width=2.73in}
   \epsfig{file=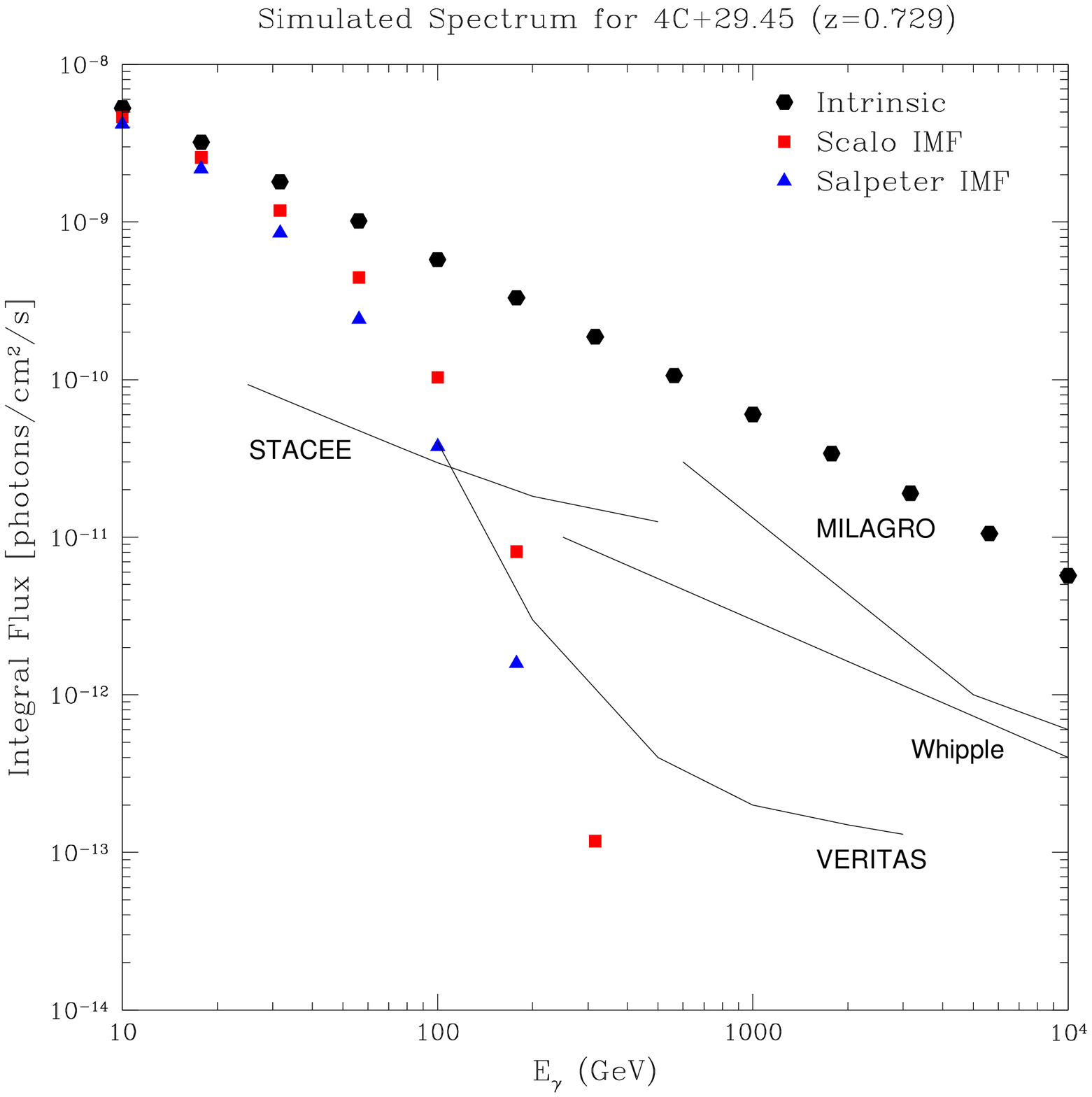,height=2.73in,width=2.73in}
   \hfill
   \epsfig{file=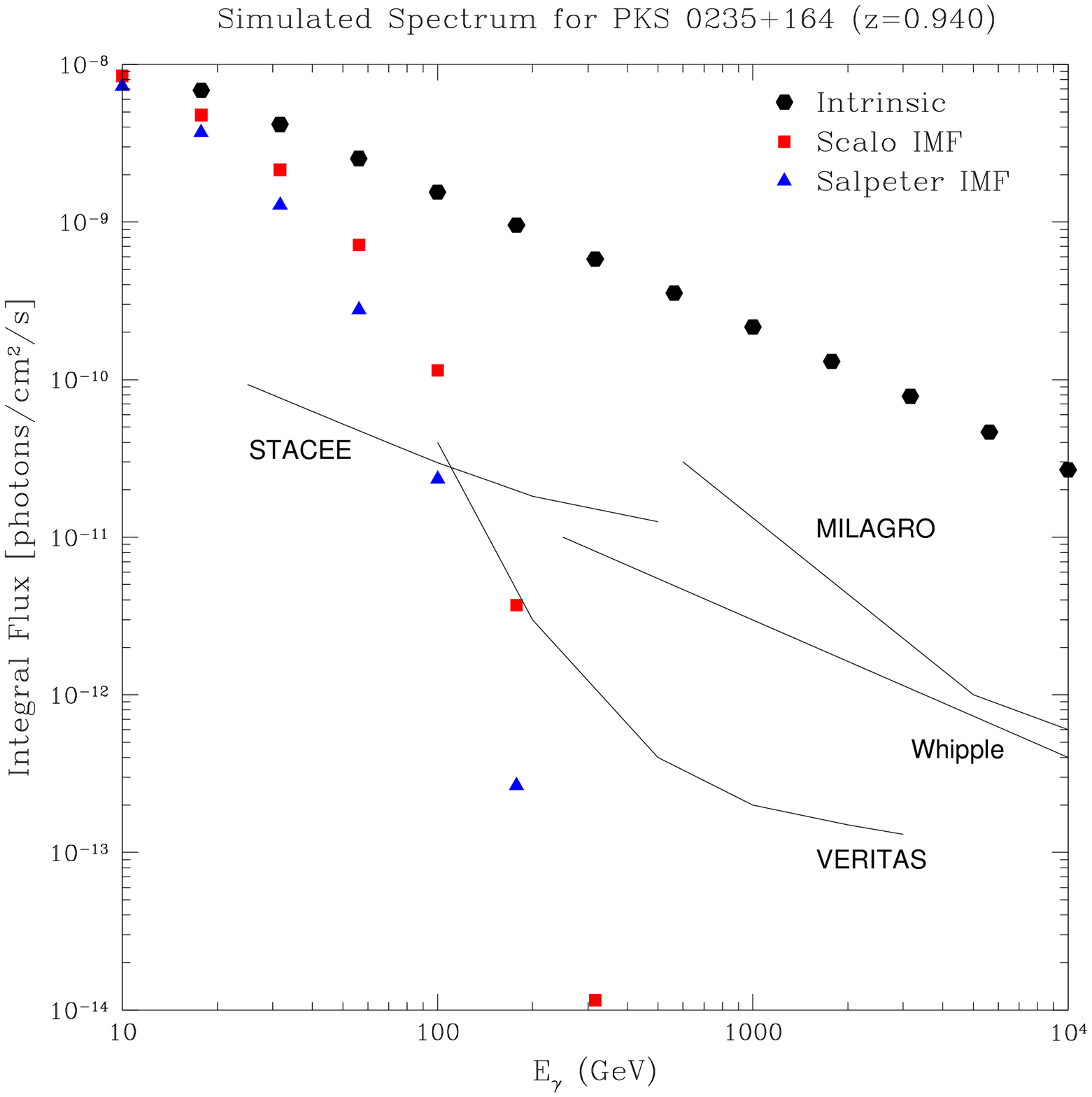,height=2.73in,width=2.73in}
\caption{Calculation of expected integral flux for six EGRET sources
   at various redshifts.}
\label{figspectra}
\end{figure}

For the absorption simulations, we assumed a simple power law for the
intrinsic spectrum of each source.  We did not assume any source
absorption effects.  Spectral indices and pre-factors for the
differential spectra were obtained from the third EGRET catalog
\cite{hartman99}, and calculations of these differential spectra followed
\cite{thompson96}.  We then modified each of these intrinsic spectra
with an absorption factor.  We assumed the functional form of this
factor to be an exponential decay, whose exponent ($\tau$) is derived
from the calculations of the EBL mentioned earlier.  Plots of $\tau$ as
a function of energy for a sample of the red-shifts considered appear in
Fig.~\ref{figtau}b.  The functional form for the simulated differential
spectrum is then:
\begin{equation}
   \dndE_{abs}\!\!\!\! = \dndE_{int}\!\!\!\!\exp[-\tau(E)]
\end{equation}
Integral fluxes were calculated numerically from this absorption
corrected differential flux.  Results are plotted in Fig.~\ref{figspectra}
for the two different stellar IMFs under consideration here.  Also
included for reference on the plots in Fig.~\ref{figspectra} are
sensitivity curves
for a representative sample of ground based gamma ray detectors.

\begin{wrapfigure}[18]{r}[0pt]{2.8in}
\epsfig{file=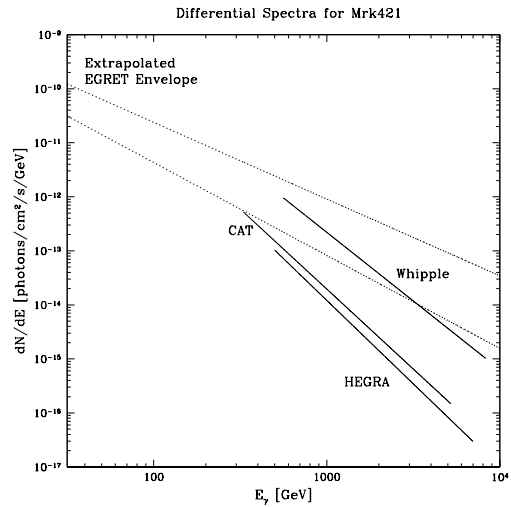,height=2.7in,width=2.7in}
\caption{TeV spectra for Mrk421 and extrapolated EGRET spectrum.}
\label{differential}
\end{wrapfigure}

Four of the six sources we considered had been previously selected for observations
by the Whipple group \cite{buckley99}, and each resulted in a non-detection.
These sources are W Comae, PKS 1604+159, 3C66A, and PKS 0235+164.  The
upper limits placed on the latter three are consistent with our model.
However, Whipple's upper limit on W Comae is significantly below both the
Scalo and Salpeter corrections to a simple EGRET extrapolated power law
spectrum.  In addition to this discrepancy for W Comae, we
are unable to reproduce Whipple's observed integral flux for Mrk 421 during
the same epoch \cite{schubnell}.  Comparison of our predicted differential
spectrum and the current observed differential spectra from Whipple, HEGRA,
and CAT \cite{aharonian99,krennrich99,piron99} also yielded disparate results,
as illustrated in
Fig.~\ref{differential}.  The poor correlation between our extrapolated
EGRET data and the TeV observations make it clear that we cannot 
yet make a strong statement about the EBL, or the corresponding
IMF, due mainly to the uncertainty of our simple power-law model of the
intrinsic source
spectrum.
In future efforts, we will exchange our simple power law model for
a more realistic simulation of the intrinsic source spectrum.  However,
while the plots in Fig.~\ref{figspectra} may not be realistic,
they do demonstrate the effects of absorption on the spectrum, and how this
feature varies with redshift.

\end{document}